\documentclass[11pt]{article}
\usepackage{epsfig}
\usepackage{graphicx}
\usepackage{rotating}

\newcommand{\BaBarYear}       {02}
\newcommand{\BaBarNumber}     {09}
\newcommand{\SLACPubNumber} {9232}

\newcommand{\BaBarType}     {CONF}  

\def\modeIV{B^{\pm} \rightarrow \pi^{\pm} \pi^{\mp} \pi^{\pm}}
\def\modeV{B^{\pm} \rightarrow K^{\pm} \pi^{\mp} \pi^{\pm}}

\def\modeXI{B^{\pm} \rightarrow K^{\pm} K^{\mp} \pi^{\pm}}
\def\modeXII{B^{\pm} \rightarrow K^{\pm} K^{\mp} K^{\pm}}

\def\modepipipi{\pi^{\pm} \pi^{\mp} \pi^{\pm}}
\def\modeKpipi{K^{\pm} \pi^{\mp} \pi^{\pm}}
\def\modeKKpi{K^{\pm} K^{\mp} \pi^{\pm}}
\def\modeKKK{K^{\pm} K^{\mp} K^{\pm}}

\newcommand{\pvec}{{\bf p}}
\newcommand{\half}{\mbox{${1\over2}$}}
\def\etal               {{\it et~al.,}}

\input pubboard/babarsym

\long\def\inst#1{\par\nobreak\kern 4pt\nobreak
    {\it #1}\par\vskip 10pt plus 3pt minus 3pt}

\begin{document}
{\pagestyle{empty}

\begin{flushright}
\babar-\BaBarType-\BaBarYear/\BaBarNumber \\
SLAC-PUB-\SLACPubNumber \\
\end{flushright}

\par\vskip 3cm

\begin{center}
\Large \bf \boldmath
Measurements of the branching fractions of charmless three-body charged $B$ decays
\end{center}
\bigskip

\begin{center}
\large The \babar\ Collaboration\\
\mbox{ }\\
May 31, 2002
\end{center}
\bigskip \bigskip

\begin{center}
\large \bf Abstract
\end{center}
We present preliminary results of searches for charged $B$ mesons 
decaying into the charmless three-body final states 
$h^{\pm}h^{\mp}h^{\pm}$, 
where $h$ = $\pi$ or $K$, 
using 51.5 \invfb\ of data collected at the \FourS\ resonance 
with the \babar\ detector at the SLAC \pep2 asymmetric $B$ Factory. 
No assumptions are made about intermediate resonances.
We measure the branching fractions
${\cal{B}}(\modeV) = (59.2 \pm 4.7 \pm 4.9) \times 10^{-6}$ and 
${\cal{B}}(\modeXII) = (34.7 \pm 2.0 \pm 1.8) \times 10^{-6}$, 
where the first error
is statistical and the second error is systematic.
In the same study, we do not observe significant signals
for the final
states $\modeIV$ and $\modeXI$, and therefore provide the 90\%
confidence upper limits
${\cal{B}}(\modeIV) < 15 \times 10^{-6}$ and 
${\cal{B}}(\modeXI) < 7 \times 10^{-6}$.
\vfill
\begin{center}{Presented at the Flavor Physics and CP Violation (FPCP) Conference, \\
$5/16 - 5/18/2002$, Philadelphia, USA}
\end{center}

\vspace{1.0cm}
\begin{center}
{\em Stanford Linear Accelerator Center, Stanford University,
Stanford, CA 94309} \\ \vspace{0.1cm}\hrule\vspace{0.1cm}
Work supported in part by Department of Energy contract DE-AC03-76SF00515.
\end{center}

\newpage

}

\begin{center}
\small

The \babar\ Collaboration,
\bigskip

B.~Aubert,
D.~Boutigny,
J.-M.~Gaillard,
A.~Hicheur,
Y.~Karyotakis,
J.~P.~Lees,
P.~Robbe,
V.~Tisserand,
A.~Zghiche
\inst{Laboratoire de Physique des Particules, F-74941 Annecy-le-Vieux, France }
A.~Palano,
A.~Pompili
\inst{Universit\`a di Bari, Dipartimento di Fisica and INFN, I-70126 Bari, Italy }
% ...

\end{center}\newpage

\setcounter{footnote}{0}

\section{Introduction}
\label{sec:Introduction}

The study of charmless hadronic $B$ decays is important to understand
the phenomenon of \CP\ violation in the Standard Model.
There has been recent theoretical progress on using three-body
decays to measure direct \CP\ violation 
and to extract the Cabibbo-Kobayashi-Maskawa angle $\gamma$~\cite{ref:GammaPaper}.
It is necessary to first observe these decays before such measurements can be made.
We present updated preliminary results on the branching fractions of
charged charmless three-body $B^{\pm} \ra h^{\pm} h^{\mp} h^{\pm}$ decays,
where $h$ = $\pi$ or $K$, with no assumptions about intermediate
resonances and with open charm contributions subtracted.
Charge conjugate initial and final states are assumed throughout
this document, unless stated otherwise.

\section{\boldmath The \babar\ detector and dataset}
\label{sec:babar}
The data used in this analysis were collected with the \babar\ detector
at the \pep2\ asymmetric \epem\ storage ring at SLAC. 
The data sample consists of 56.2 million \BB\ pairs, corresponding to an
integrated luminosity of 51.5~\invfb\ collected at the \FourS\ resonance (on-resonance) during
the 2000-2001 run.
In addition, a total integrated luminosity of 6.4~\invfb\ was taken at 40~\mev\ below the
\FourS\ resonance (off-resonance), and was used to characterise the backgrounds
from \epem\ annihilation into light \qqbar\ pairs. 

The \babar\ detector is described in detail elsewhere~\cite{ref:babar}; 
the main parts relevant for the analysis of three charged particle final 
states are the tracking and particle identification sub-detectors.

The 5-layer double-sided silicon vertex tracker (SVT)
measures the impact parameters, angles, and transverse momenta of tracks down to 65~\mevc.
Outside the SVT is a 40-layer drift chamber (DCH) that measures the transverse momenta
of tracks from their curvature in the 1.5-T solenoidal magnetic field.
The SVT and DCH also are used to determine the mean
ionisation energy loss of tracks to help identify
charged particles. The tracking system  
has a momentum resolution of 0.5\% for a transverse momentum of 1.0~\gevc.

Surrounding the DCH is a detector of internally reflected Cherenkov 
radiation (DIRC), which provides charged hadron identification in the barrel region. 
Charged particles are identified by the Cherenkov angle $\theta_c$
and the number of photons measured with the DIRC.
The typical separation between pions and kaons varies from $> 8~\sigma$ at 2.0\gevc to 2.5$\sigma$
at 4.0\gevc, where $\sigma$ is the average resolution on $\theta_c$.
The kaon selection efficiency is approximately 80\%, which is the product of 
the particle identification algorithm efficiency with
geometrical acceptance, for a pion mis-identification probability of 2\%.

The DIRC is surrounded by an electromagnetic
calorimeter (EMC), made up of 6580 CsI(Tl) crystals, which is used
to measure the 
energies and angular positions of photons and electrons with excellent 
resolution.
The EMC is used to veto electrons in this analysis; 
the probability of mis-identifying electrons as pions is approximately
5\%, while the probability of mis-identifying pions as electrons is
below 0.3\%.

\section{\boldmath Analysis method}
\label{sec:Analysis}
The total branching fraction for each $B^{\pm} \ra h^{\pm} h^{\mp} h^{\pm}$
mode is measured over the whole Dalitz plot - all resonant and 
non-resonant contributions are included. 
A set of selection criteria is applied to reconstruct each mode separately.
Each Dalitz plot is divided into many equal area cells to enable us to find the 
selection efficiency as a function of position in the Dalitz plot. 
We also take into account
continuum backgrounds and cross-feed between each signal mode from $K$ 
and $\pi$ mis-identification. 

\subsection{Candidate Selection}

We reconstruct $B$ candidates from charged tracks, where each track must have at least 
12 hits in the DCH, a maximum momentum of 10~\gevc, a minimum transverse 
momentum of 100~\mevc, and must originate from the beam-spot. We find
three-charged track combinations to form the $B$ candidates, and we 
require that their energies
and momenta satisfy kinematic constraints appropriate for $B$ mesons.

There are two variables we use for this, the first of which is 
the beam-energy substituted mass $\mes = \sqrt{(E^2_b - \pvec^2_B)}$.
The energy of the $B$ candidate is defined as 
$E_b = (\half s + \pvec_0 \cdot \pvec_B)/E_0$, where $\sqrt{s}$ and $E_0$
are the total energies of the \epem\ system in the centre-of-mass (CM)
and laboratory frames, respectively, and $\pvec_0$ and $\pvec_B$
are the momentum vectors in the laboratory frame of the \epem\
system and the $B$ candidate, respectively. The $\mes$ value should be close
to the nominal $B$ mass for signal events.

The second variable we use is the 
energy difference between the reconstructed $B$ candidate energy and 
the beam energy, 
$\DeltaE = E_B^* - \sqrt{s}/2$, where $E_B^*$ is the energy of the 
$B$ candidate in the CM system. For this analysis, we assume the appropriate
mass hypothesis for each charged track in a given decay mode under study in calculating \DeltaE.
For signal events, $\Delta E$
should be centred at zero.
The typical $\Delta E$ separation between modes that
differ by substituting a kaon for a pion in the final state is $45~\mev$.

We use \dedx information from the SVT and DCH, 
and the Cherenkov angle and number of
photons measured by the DIRC for tracks with momenta above 700~\mevc, 
to identify charged pions and kaons. Kaons are selected with
requirements made to the product of the likelihood ratios determined from these measurements.
The likelihood ratio requirements are established by requiring the 
probability of mis-identifying pions as kaons be
below 5\%, up to a momentum of 4.0~\gevc. Pions
are required to fail the kaon selection.
We veto
electron candidates by requiring that they fail a selection based on information from
\dedx, shower shapes in the EMC and the ratio of the shower energy and track momentum.

Since we are only interested in charmless decays, 
we need to veto candidates that contain charm mesons.
We remove $B$ candidates when the 
invariant mass of the combination of any two of its daughter tracks (of opposite charge) is
within $3~\sigma$ of the mass of $D^0$, $J/\psi$ or $\psi(2S)$ mesons. 
Here, $\sigma$ is 10.0~\mevcc for $D^0$ and 15.0~\mevcc for $J/\psi$ and $\psi(2S)$. 
All possible kaon and pion combinations are tested for the $D^0$ veto, while only
the $K^+K^-$ and $\pi^+\pi^-$ hypotheses are tested 
for the $J/\psi$ and $\psi(2S)$ vetoes, since
the background from these decays is from leptonic decays, in which the leptons
have been mis-identified as pions or kaons. 
The electron veto helps to reduce the combinatorial background from $J/\psi$ and $\psi(2S)$ decays
that would otherwise pass the $3~\sigma$ invariant mass veto.

\subsection{Background Suppression and Characterisation}

In addition to these candidate selection requirements, we need to suppress
backgrounds from light quark and charm continuum production. We reduce these by imposing
requirements on two topological event shape variables computed in the \FourS\ rest frame. 

The first event shape variable
is the cosine of the angle $\theta_T$ between the thrust axis of the selected $B$ candidate
and the thrust axis of the rest of the event, i.e. all charged tracks and neutral particles
not in the $B$ meson candidate. For continuum backgrounds, the directions of the two axes tend to
be aligned because the daughters of the reconstructed candidate generally lie along the dijet axis
of such events.
Therefore, the distribution of $|$cos$\theta_T|$ is strongly peaked 
towards unity.
For $B$ events, the distribution of $|$cos$\theta_T|$ is isotropic because 
the decay products from the two $B$ mesons are essentially 
independent of each other.
The difference in the $|$cos$\theta_T|$ dependence 
allows us to discriminate between signal $B$ decays and continuum background.

The second event shape variable is a Fisher
discriminant~\cite{ref:Fisher} ${\cal{F}}$, which is formed from 
the linear 
combination
\begin{equation}
{\cal{F}} = \sum_{i=1}^9 \alpha_i x_i
\end{equation}
of the input variables $x_i$. The available variables are
the summed scalar momenta of all charged and neutral particles from the rest of the event
within nine nested cones coaxial with the thrust axis of the $B$ candidate.
The coefficients $\alpha_i$
are chosen to maximise the separation between signal and background events. They
are calculated for each signal mode separately using
Monte Carlo signal and light quark continuum events.

\begin{figure}[!htb]
\begin{center}
\hspace*{-1.0cm}
\mbox{\epsfig{file=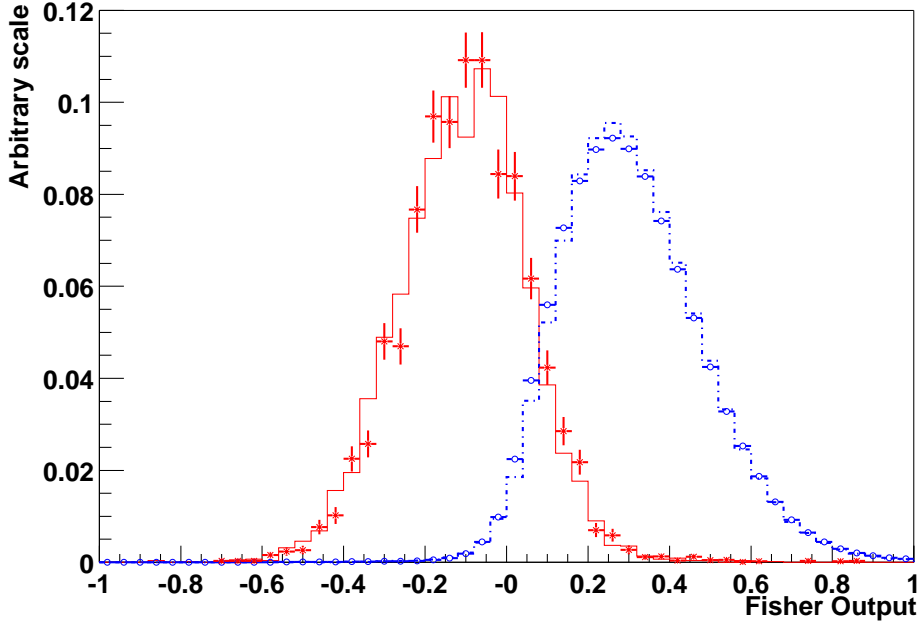,width=9.0cm,angle=-90}}
\end{center}
\caption[Distribution of the Fisher discriminant for $B^- \ra D^0 \pi- (D^0 \ra K^- \pi^+$) 
and background events.]
{Normalised distributions of the Fisher discriminant
for $B^- \ra D^0 \pi^-, D^0 \ra K^- \pi^+$ Monte Carlo events 
(solid histogram), on-resonance $D^0 \pi^-$ data (solid points), 
light quark continuum Monte Carlo events (dotted histogram)
and off-resonance data (dotted points).}
\label{fig:Fisher}
\end{figure}

Figure~\ref{fig:Fisher} shows the Fisher distributions for $\FourS$ events, 
from the control sample $B^- \ra D^0 \pi^-, D^0 \ra K^- \pi^+$ 
in Monte Carlo simulation and on-resonance data, 
and for background, from off-resonance data and 
light-quark continuum Monte Carlo events.

The selection criteria for the event shape variables, shown in 
Table~\ref{tab:BGcuts},
is optimised separately for each signal mode to achieve maximum
sensitivity for the branching fraction.
\begin{table}[htb!]
\caption{Selection requirements on the event shape variables for each
signal mode.}
\begin{center}
{\small
\begin{tabular}{|l||c|c|}
\hline
Signal Mode & $|$cos$\theta_T|$ & ${\cal{F}}$ \\
\hline
\hline
$\modepipipi$ & $< 0.575$ & $< -0.11$ \\
\hline
$\modeKpipi$ & $< 0.700$ & $< -0.03$ \\
\hline
$\modeKKpi$ & $< 0.725$ & $< 0.10$ \\
\hline
$\modeKKK$ & $< 0.875$ & $< 0.30$ \\
\hline
\end{tabular}
}
\end{center}
\label{tab:BGcuts}
\end{table}

Despite using the powerful event shape variables mentioned above, there are still significant
backgrounds that must be explicitly subtracted to extract a signal. The residual 
background level is
estimated from the observed number of events in a sideband region, located
near to the signal region in the \mes- \DeltaE\ plane, 
and then extrapolating into the
signal region by using a multiplicative factor, $R$. 
We define $R$ to be the ratio of the number of background candidates
in the signal region to the number in the sideband region. 
In order to determine $R$, the shape of the background
distribution as a function of $\mes$ is parameterised according to the 
ARGUS function~\cite{ref:Argus},
and is measured using the upper sideband in the \DeltaE variable 
in on-resonance data ($0.1 < |\DeltaE| < 0.25\gev$). 
A quadratic function is used to parameterise the background distribution
as a function of \DeltaE. 
The ratio of the areas under the shape function in $\Delta E$ and
$\mes$ in the signal and sideband regions is equal to $R$. The 
uncertainty of the value of $R$ is dominated by the uncertainty
of the shape parameter for the ARGUS function.
Off-resonance data give a consistent value of $R$.

\subsection{Branching Fraction Calculation}

As mentioned previously, the branching fractions for each signal
mode are measured over the whole Dalitz plot, and each Dalitz plot is 
divided up into many cells so that the bin-by-bin 
variation of the selection efficiency 
can be found for each plot.

Taking $\epsilon_i$ to be the Monte Carlo efficiency of 
reconstructing the signal in the
$i^{th}$ bin in the Dalitz plot, the 
branching fraction for the signal mode is given by:
\begin{equation}
{\cal{B}} = \frac{1} {N_{\BB}} \sum_i \frac{(N_{1i} - R N_{2i})} {\epsilon_i} = 
\frac{1} {N_{\BB}} \sum_i \frac{S_i}{\epsilon_i},
\label{eqn:firstBR}
\end{equation}
where $N_{\BB}$ is the total number of \BB\ pairs, $R$ is the background 
extrapolation
factor into the signal region, and $N_{1i}$ and $N_{2i}$ are 
the number of events observed in the signal and grand sideband (GSB) regions, 
respectively, 
for the $i^{\mbox{th}}$ Dalitz plot bin. No significant differences were 
found for the value of $R$ in different regions of the Dalitz plot, 
so an average value is used for all bins.
$S_i$ is the number of background
subtracted signal events for the $i^{\mbox{th}}$ Dalitz plot bin.

The signal region is defined to be 
$|\mes - m_{B}| < 8.0~\mevcc$ and 
$|\DeltaE - \left<\DeltaE\right>| < 60.0~\mev$, 
where $\left<\DeltaE\right>$ is the mean
value of $\DeltaE$ for on-resonance data for the calibration sample
$B^- \ra D^0 \pi^-, D^0 \ra K^- \pi^+$, and $m_{B}$ 
is the nominal mass of the charged
$B$ meson~\cite{ref:pdg2000}. 
The GSB region is defined to be
$5.21 < \mes < 5.25~\gevcc$ and 
$|\DeltaE - \left<\DeltaE\right>| < 100.0~\mev$.

The probability of a kaon being mis-identified as a pion
is 20\%, which includes the efficiency of the particle identification algorithm
and the geometrical acceptance. 
This means that there is significant cross-feed into the signal region 
from the decay mode that has one more kaon, which must be subtracted
for each bin, $i$. To see how this is done, consider the branching fraction calculation
for $\modeXII$. 
This channel has negligible cross-feed from the other modes, so we have
\begin{equation}
{\cal{B}}({\modeXII}) = \frac{1} {N_{\BB}} \sum_i \frac{S_i} {\epsilon_i} = 
\frac{N_{3K}} {N_{\BB}},
\label{eqn:BR3K}
\end{equation}
where $N_{3K}$ is the total number of $\modeXII$ events in the data.
The channel $\modeKKpi$, on the other hand,
has significant cross-feed from $\modeKKK$, and this must be subtracted from the signal. The 
branching fraction for this mode is given by
\begin{equation}
{\cal{B}}({\modeXI}) = \frac{1} {N_{\BB}} \sum_i \frac{(S_i - N_{3K} \epsilon''_i)} {\epsilon_i} 
= \frac{N_{KK \pi}} {N_{\BB}},
\label{eqn:BRKKpi}
\end{equation}
where $N_{KK \pi}$ is the total number of $\modeXI$ events in the data, and $S_i$ and
$\epsilon_i$ refer to those quantities for $\modeXI$. Here, $\epsilon''_i$ is 
the probability for reconstructing $\modeXII$ events using the selection criteria for $\modeXI$. 
It is determined from Monte Carlo simulation 
by generating events uniformly in phase space and 
determining the cross-feed selection efficiency in each Dalitz plot bin.

The branching fraction for $\modeV$ is given by
\begin{equation}
{\cal{B}}({\modeV}) = \frac{1} {N_{\BB}} \sum_i \frac{(S_i - N_{KK \pi} \epsilon''_i - n_{Di})} {\epsilon_i} = 
\frac{N_{K \pi \pi}} {N_{\BB}},
\label{eqn:BRKpipi}
\end{equation}
where $N_{K\pi\pi}$ is the total number of $\modeV$ events in the data, 
and $S_i$ and
$\epsilon_i$ refer to the number of background subtracted
events in the signal region and selection efficiency for $\modeV$, 
respectively. Here,
$\epsilon''_i$ is the probability for reconstructing $\modeXI$ events 
using the selection criteria for $\modeV$.
This channel has some
$D^0$ contamination from candidates falling outside
the $3~\sigma$ invariant mass veto, which must be subtracted.
This is represented by the term $n_{Di}$, which
is the number of $D^0$ events that is expected to populate the 
$i^{\mbox{th}}$ bin in the
Dalitz plot. This is estimated by finding the probability of reconstructing 
$D^0$ Monte Carlo events, for each bin $i$, 
using the selection criteria for $\modeV$, and multiplying this 
by the measured
branching fraction for the $D^0$ mode~\cite{ref:pdg2000} 
and the total number of \BB\ pairs,
$N_{\BB}$. 
The total expected number of $B^- \ra D^0 \pi^-, D^0 \ra K^- \pi^+$ 
events that must be subtracted across the full Dalitz plot 
for this channel is $47 \pm 8$. 
The values of $n_{Di}$ are non-zero only for bins close to the $D^0$ 
resonance bands.

Finally, the branching fraction for $\modeIV$ is given by
\begin{equation}
{\cal{B}}({\modeIV}) = \frac{1} {N_{\BB}} \sum_i \frac{(S_i - N_{K \pi \pi} \epsilon''_i - n_{Di})} {\epsilon_i} = 
\frac{N_{3 \pi}} {N_{\BB}},
\label{eqn:BR3pi}
\end{equation}
where $N_{3 \pi}$ is the total number of $\modeIV$ events in the data,
$\epsilon''_i$ is the probability for reconstructing $\modeV$ events 
using the selection
criteria for $\modeIV$, and $S_i$ and
$\epsilon_i$ refer to the number of background subtracted
events in the signal region and selection efficiency for $\modeIV$, 
respectively.
Again, there are
some $B^- \ra D^0 \pi^-, D^0 \ra K^- \pi^+$ events ($23 \pm 5$) 
that pass the selection criteria for this channel, because the kaon from
the $D^0$ decay is mis-identified as a pion, 
and where the invariant mass of
the $D^0$ meson lies outside the $3~\sigma$ invariant mass window.
This background is subtracted from the signal.

In addition to the cross-feed 
where only one of the kaon tracks
is mis-identified as a pion,
there can also be cross-feed where either two kaons are mis-identified 
as pions 
(probability of 4\%) or when one of the pions is mis-identified as a kaon
(probability of 2\%).
These are smaller, second-order cross-feed effects, and so it is 
adequate to
subtract the average number of events over the whole Dalitz plot. 
If $n_x$ is the average
number of second-order cross-feed events that has to be subtracted (i.e. the number of
events reconstructed divided by the appropriate cross-feed efficiency),
then we finally have
\begin{equation}
{\cal{B}} = \frac{1} {N_{\BB}} \left(\sum_i \frac{(S_i - N_x \epsilon''_i - n_{Di})} {\epsilon_i} - n_x \right),
\label{eqn:fullBR}
\end{equation}
where $N_x$ is the total number of events from the channel that contributes to most of the
(first-order) cross-feed, e.g. $N_x = N_{K\pi\pi}$ for $\modeIV$. 
\begin{table}[!hbt]
\caption
{Efficiencies and cross-contamination probabilities between the signal modes derived from
Monte Carlo samples. For example, the probability that an event $\modeKpipi$ will be reconstructed
as $\modepipipi$ is $(1.7 \pm 0.1) \times 10^{-2}$.}
\begin{center}
{\small
\begin{tabular}{|l||c|c|c|c|} 
\hline
 & \multicolumn{4}{|c|}{Input Decay Mode}\\
Selection Criteria & \multicolumn{4}{|c|}{}\\ 
\cline{2-5}
Hypothesis & $\modepipipi$ & $\modeKpipi$ & $\modeKKpi$ & $\modeKKK$\\
\hline
$\modepipipi$& $(15.3 \pm  0.2) \times 10^{-2}$ & $(1.7 \pm 0.1) \times 10^{-2}$ & $(1.4 \pm 0.9) \times 10^{-4}$ & $(1.1 \pm 3.2) \times 10^{-5}$\\
$\modeKpipi$& $(3.6 \pm 0.4) \times 10^{-3}$ & $(15.1 \pm  0.2) \times 10^{-2}$ & $(3.2 \pm 0.2) \times 10^{-2}$ & $(4.0 \pm 1.7) \times 10^{-4}$\\
$\modeKKpi$& $(0.0 \pm 0.2) \times 10^{-3}$ & $(2.9 \pm 0.4) \times 10^{-3}$ & $(17.7 \pm  0.3) \times 10^{-2}$ & $(5.5 \pm 0.2) \times 10^{-2}$\\
$\modeKKK$ & $(0.0 \pm 0.2) \times 10^{-3}$ & $(0.0 \pm 0.2) \times 10^{-3}$ & $(1.7 \pm 0.2) \times 10^{-3}$ & $(21.6 \pm  0.3) \times 10^{-2}$\\
\hline 
\end{tabular}

}
\end{center}
\label{tab:EfficiencyMatrix}
\end{table}

The Dalitz plot for each signal mode is divided into cells with equal 
area 1.0~$\gev^4$, and
large samples of Monte Carlo signal events are used to obtain the 
signal and cross-feed selection efficiencies across each Dalitz plot. 
Table~\ref{tab:EfficiencyMatrix} shows the signal and cross-feed 
selection efficiencies
for the modes, averaged over the Dalitz plots.

\section{\boldmath Experimental Uncertainties}
\label{sec:Systematics}
There are several sources of uncertainty for the branching fraction measurements, 
which come from the various terms in Eq.~\ref{eqn:fullBR}. 
The statistical uncertainties come from the
number of events observed in the signal and GSB regions, $N_1$ and $N_2$.
The factor $R$, found independently for each mode, has a systematic uncertainty arising from 
the uncertainty in the
fitted ARGUS shape parameter, $\xi$. 
The main sources of $B$-related background are $D^0$ decays (the $n_{Di}$ term) 
and cross-feed from the other signal modes (the $N_x$ and $n_x$ terms), 
owing to kaon and pion mis-identification. We deal with these by explicitly subtracting 
them from the signal. The uncertainty in the number of $D^0$ events that have to be subtracted
comes from the uncertainty on the published measured branching fractions~\cite{ref:pdg2000}, the
number of $\BB$ events (mentioned below) and the
selection efficiency.

Since there are a lot of terms used to calculate the branching fraction, it
is worthwhile to go through what uncertainties they contribute to the end result.
If we let $X_i$ represent the term within parenthesis in Eq.~\ref{eqn:fullBR}, then the 
fractional uncertainty on the branching fraction is
\begin{equation}
\left( \frac{\Delta {\cal{B}}}{{\cal{B}}} \right)^2 = 
\left(\frac{\Delta N_{\BB}}{N_{\BB}}\right)^2 +
\left(\frac{\Delta (\sum_i X_i)} {(\sum_i X_i)}\right)^2
+ \delta_{e}^2,
\label{eqn:BRFracErr}
\end{equation}
where $\delta_{e}$ is the fractional systematic error for the efficiency
that comes from differences between Monte Carlo simulation and on-resonance data,
shown in Table~\ref{tab:effsyst}.
Going through the terms for $X_i$, we have
\clearpage
\begin{eqnarray}
\left(\Delta (\sum_i X_i)\right)^2 & = & \sum_i \frac{N_{1i}}{\epsilon_i^2} + 
R^2 \sum_i \frac{N_{2i}}{\epsilon_i^2} + 
(\Delta R)^2\left(\sum_i \frac{N_{2i}}{\epsilon_i}\right)^2 \nonumber \\
& & + N_x^2 \sum_i \left( \frac{\Delta \epsilon''_i}{\epsilon_i}\right)^2 +
(\Delta N_x)^2 \left( \sum_i \frac{\epsilon''_i}{\epsilon_i}\right)^2 + 
\sum_i \left( \frac{\Delta n_{Di}}{\epsilon_i}\right)^2 \nonumber \\
& & + \sum_i \left( \Delta \epsilon_i \frac{N_{1i} - RN_{2i} - N_x \epsilon''_i - n_{Di}}{\epsilon_i^2}\right)^2 + (\Delta n_x)^2.
\label{eqn:xErrEqn}
\end{eqnarray}
The first two terms on the right hand side (R.H.S.) 
are the statistical uncertainties on
the number of events in the signal and GSB regions, while the third term
represents the systematic variation for the background extrapolation
factor, $R$. The uncertainties from the cross-feed subtraction are
represented by the next three terms, while the penultimate term is that
for the bin-by-bin uncertainty for the efficiency. The last term is the
uncertainty for the number of second-order cross-feed events.
The various sources of error mentioned above will be shown in detail 
for each branching fraction result.

The uncertainties on the signal efficiencies and cross-feed probabilities 
are the combination of statistical errors
on the number of events selected in the Monte Carlo samples relative to the 
total number generated, as well as systematic uncertainties arising from the difference
between Monte Carlo simulation and on-resonance data.

The average fractional Monte Carlo statistical uncertainties of the signal efficiencies per Dalitz plot bin ($\Delta \epsilon_i/\epsilon_i$) are 
7.0\% for $\modepipipi$, 9.1\% for $\modeKpipi$, 9.5\% for $\modeKKpi$ and 7.4\% for $\modeKKK$.

The Monte Carlo simulation is subject to systematic uncertainties from tracking and
particle identification efficiencies.
Residual differences in the tracking selection efficiencies between 
on-resonance data and Monte Carlo
simulation contributes a fractional uncertainty of 0.8\% on the efficiency per track. 
This uncertainty is added
coherently for all three tracks used to reconstruct each $B$ meson. Both the electron veto and kaon selections 
have fractional systematic uncertainties of 1.0\%. These uncertainties are added coherently.

Possible differences between the behaviour of Monte Carlo simulation 
and on-resonance data are also examined
for the
Fisher distributions used to discriminate signal $B$ decays from light quark continuum events. 
The control samples $B^- \ra D^0 h^-, D^0 \ra h^- h^+$, 
where $h$ = $\pi$ and/or $K$, 
which have similar kinematics to the signal modes, are used
to compare the signal Fisher distributions between on-resonance 
and Monte Carlo data,
using the Fisher coefficients derived from Monte Carlo signal and 
\qqbar samples.
The choice for $h$ is made such that
the final state of the control sample decay has the same number of kaons and pions as those for 
the signal mode.

The Fisher distribution for off-resonance data agrees with that for 
light quark continuum Monte Carlo events, which can be seen in Fig.~\ref{fig:Fisher}.
There are very slight differences between the Fisher distributions 
for Monte Carlo signal events and on-resonance data. To quantify this difference, 
the Fisher distributions are fitted to double Gaussian functions. 
The differences in the mean and width values
between on-resonance and Monte Carlo $D^0 h$ events are used to shift and scale the Fisher 
distributions for the signal Monte Carlo modes. 
The change in the selection efficiency gives an estimate
of the correction factor necessary for the requirement on the Fisher variable used in the selection 
for each mode, which is found to be very small (approximately 1\%). 
The systematic uncertainty on this correction is found by varying the
parameters of the Fisher distribution for the signal Mode Carlo modes by the (scaled) uncertainties found
for the fitted parameters for the $D^0$ control sample.

The resolutions for $\DeltaE$ and $\mes$ in data differ by a negligible amount from Monte Carlo
predictions. The main source of uncertainty arises from a $+7~\mev$ shift in the mean value
of $\DeltaE$ observed in the control sample $B^- \ra D^0 \pi^-, D^0 \ra K^- \pi^+$ in on-resonance data, 
which we correct for in the Monte Carlo. This contributes a fractional systematic 
uncertainty of 1\%.

Table~\ref{tab:effsyst} gives a summary of the 
systematic uncertainties to the efficiency for each mode 
(not including the Dalitz plot variation).
The fractional uncertainties
for the Dalitz plot variation for the cross-feed probabilities 
($\Delta \epsilon''_i/\epsilon''_i$) are approximately 30\%.

\begin{table}[htb!]
\caption{Fractional systematic uncertainties for the Monte Carlo 
efficiency for each signal mode. 
The uncertainties are added in quadrature in the total.}
\begin{center}
{\small
\begin{tabular}{|l||c|c|c|c|}
\hline
Source of Uncertainty & \multicolumn{4}{|c|}{Fractional error on efficiency (\%)} \\
\cline{2-5}
& $\modepipipi$   & $\modeKpipi$ & $\modeKKpi$ & $\modeKKK$ \\
\hline
\hline
Tracking & 2.4 & 2.4 & 2.4 & 2.4 \\
\hline
Fisher Discriminant & 2.1 & 0.7 & 0.5 & 1.7 \\
\hline
Particle Identification & 6.0 & 5.0 & 4.0 & 3.0 \\
\hline
$\DeltaE$ and $\mes$ & 1.0 & 1.0 & 1.0 & 1.0 \\
\hline
\hline
Total    & 6.8 & 5.6 & 4.7 & 4.2 \\
\hline
\end{tabular}
}
\end{center}
\label{tab:effsyst}
\end{table}

Finally, there is a systematic uncertainty on the overall normalisation, $N_{\BB}$, 
which is obtained from
a dedicated study to find the number of $B$ mesons produced in the data sample. 
This is found to have a systematic uncertainty of 1.5\%.

\section{\boldmath Physics results}
\label{sec:Physics}
Our preliminary measurements of the branching fractions for the signal modes are summarised in 
Table~\ref{tab:DataDPResults}.
The top few rows of this table show the total number of events observed in the 
signal and GSB regions, as well as the average signal reconstruction
efficiencies for each mode and the values of the background extrapolation factor $R$.

The row labelled 1) shows the sum over Dalitz plot bins 
of the number of events observed in the signal
region divided by the signal efficiency. The error on these quantities 
is the first error shown in Eq.~\ref{eqn:xErrEqn}, and only includes the uncertainty in
the number of signal events, $N_{1i}$. 

The next row, labelled 2), shows the sum over
Dalitz plot bins of the expected number of background events divided by the
signal efficiency. The errors shown for these values are the second and third 
terms on the R.H.S. of Eq.~\ref{eqn:xErrEqn}, respectively. They correspond to the
statistical uncertainty in $N_{2i}$, and the systematic error for $R$, which is
dominant.

Row 3) shows the
expected number of cross-feed events (from $K/\pi$ mis-identification). The errors
on these quantities represent the fourth and fifth terms on the R.H.S. of 
Eq.~\ref{eqn:xErrEqn} only. Note that
the $\modeXII$ mode does not have a cross-feed term, since this is negligible.

The expected number of $D^0$ events passing the
selection criteria for $\modeIV$ and $\modeV$ 
is shown in row 4), where the error for each value is
the sixth term on the R.H.S. of Eq.~\ref{eqn:xErrEqn}. 

The second-order cross-feed
terms, $n_x$, are shown in row 5). There are only entries for 
$\modeV$, from $\modeXII$ events, and for $\modeXI$, from $\modeV$ cross-feed. 
The errors for these values are dominated
by the uncertainties in the second-order cross-feed probabilities.
Note that the $n_x$ term for $\modeV$ is negative, which
compensates for the extra cross-feed
background of $\modeXII$ events that is mis-identified as $\modeXI$, and then
in turn passes the selection criteria for $\modeV$.

The various contributions to the signal and background terms are shown
in row 6). The first uncertainties are the combination of the statistical
errors for the number of events in the signal and GSB regions - they correspond
to the sum in quadrature of the error for row 1), and the first error in row 2).
The second error for the entries in row 6) corresponds to the quadrature
sum of all the other systematic errors from rows 2) to 5). The third error
for row 6) is that from the penultimate term on the R.H.S. of Eq.~\ref{eqn:xErrEqn},
i.e. the uncertainty for the selection efficiency for each of 
the Dalitz plot bins separately (not the average).
The last error for row 6) is just the fractional systematic uncertainty
for the efficiency correction factors, given in Table~\ref{tab:effsyst}.

The last row in Table~\ref{tab:DataDPResults}
shows the branching fraction results, where the 
first uncertainties are the statistical errors
on the number of signal and background events, 
while the second uncertainties are the sum in quadrature of all systematic errors.
The dominant systematic uncertainty for $\modeXII$ is 
the systematic correction factor for the efficiency, while for $\modeIV$ and $\modeXI$,
the background extrapolation factor $R$ dominates. For $\modeV$, both
uncertainties contribute equally to the systematic error.

As a consistency check, the branching fraction for the control sample
$B^- \ra D^0 \pi^-, D^0 \ra K^- \pi^+$ is measured to be
$(180 \pm 4 \pm 11) \times 10^{-6}$, which agrees with the 
previously measured value of $(203 \pm 20) \times 10^{-6}$~\cite{ref:pdg2000}.

\begin{figure}[!htb]
\begin{center}
\hspace*{-1.0cm}
\mbox{\epsfig{file=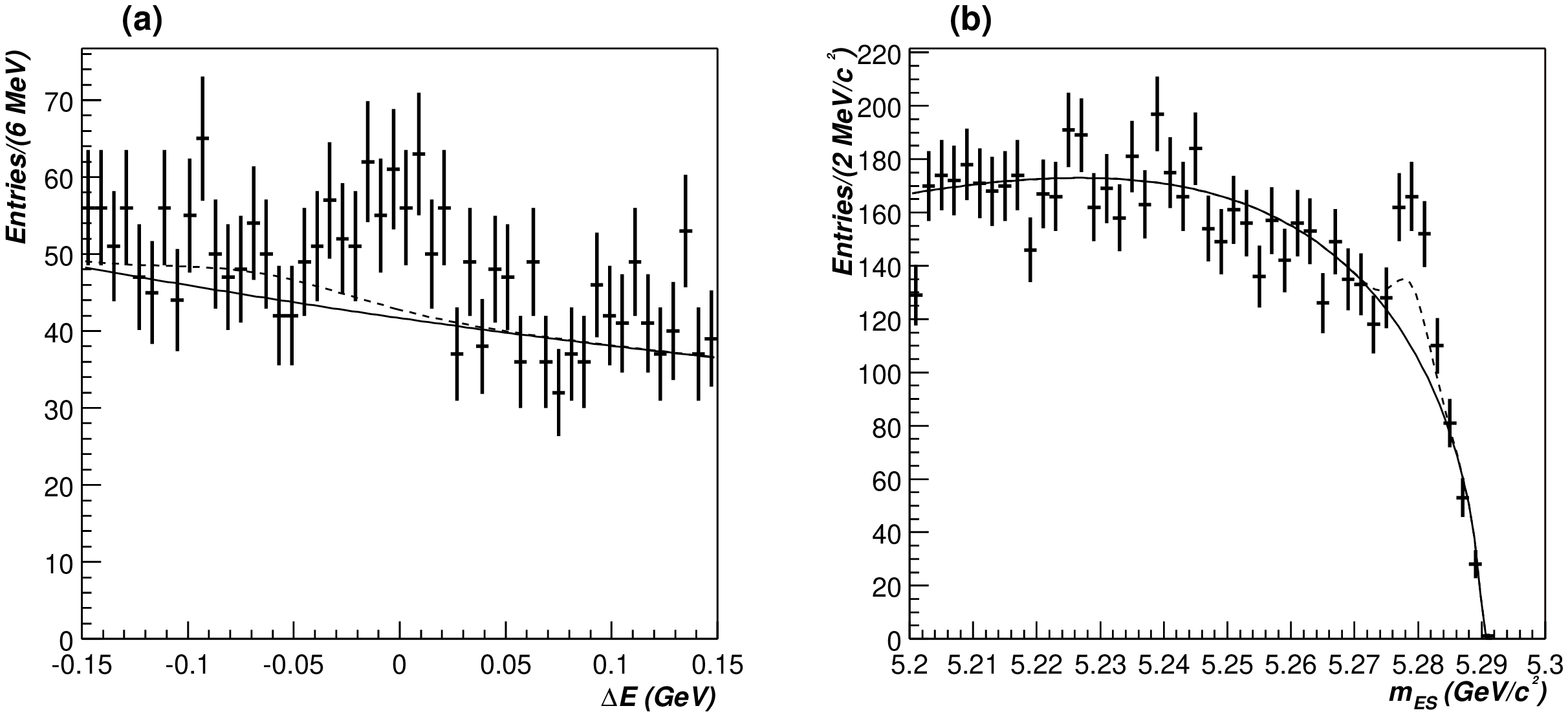,width=7.25cm,angle=-90}}
\end{center}
\caption[On-resonance signal region $\DeltaE$ and $\mes$ distributions for $\modeIV$]
{On-resonance signal region $\DeltaE$ (a) and $\mes$ (b) 
distributions for $\modeIV$. The solid lines show the expected level of continuum background,
using appropriately normalised background shapes from the sideband regions
in on-resonance data.
The dotted lines show the expected level of
\BB\ background, which is obtained from the sum of
Gaussian distributions from Monte Carlo estimated
cross-feed and $D^0 \pi$ events, each 
normalised to the number of events observed in on-resonance data 
that passed the selection criteria for $\modeIV$.}
\label{B3piCut0K_dEandmESData}
\end{figure}

\begin{figure}[!htb]
\begin{center}
\hspace*{-1.0cm}
\mbox{\epsfig{file=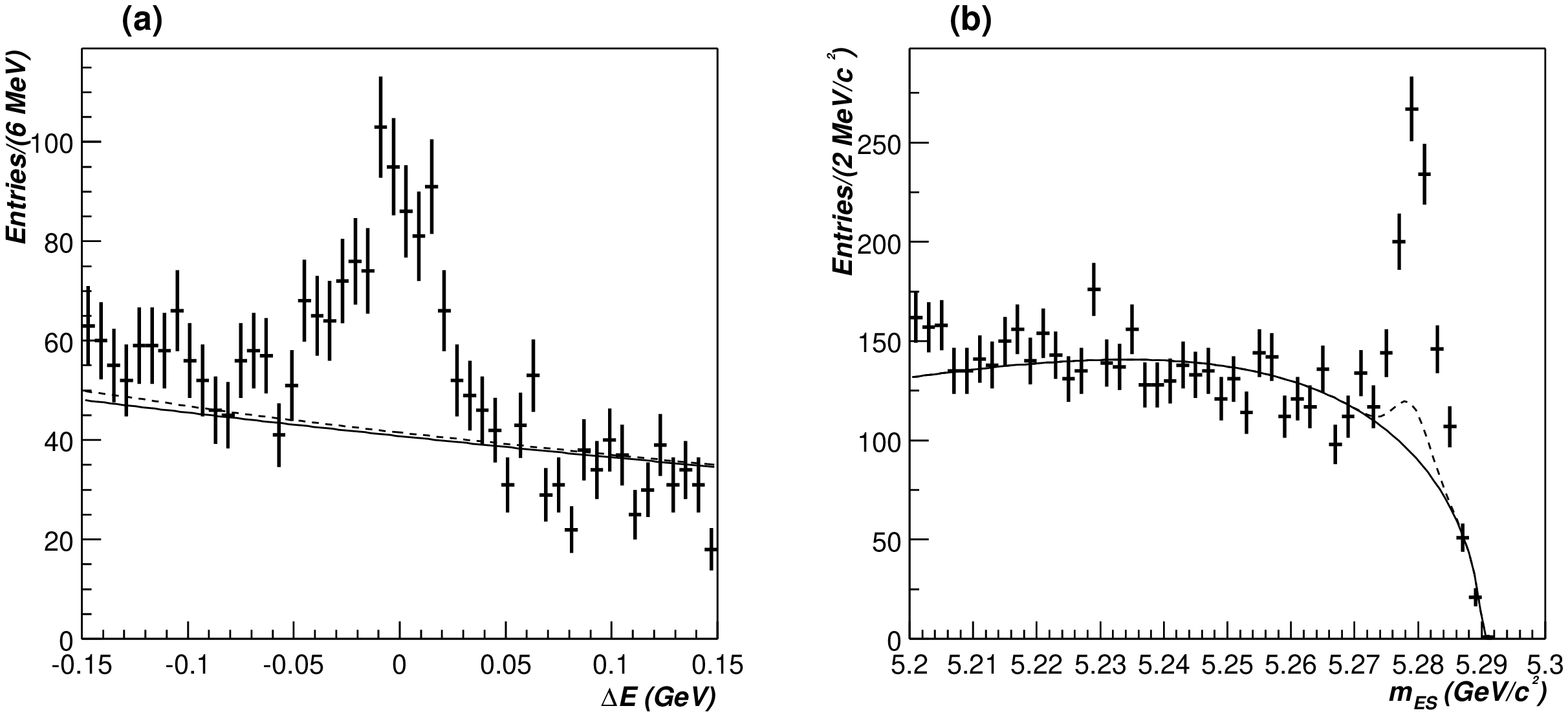,width=7.25cm,angle=-90}}
\end{center}
\caption[On-resonance signal region $\DeltaE$ and $\mes$ distributions for $\modeV$]
{On-resonance signal region $\DeltaE$ (a) and $\mes$ (b) distributions
for $\modeV$. The solid lines show the expected level of continuum background,
using appropriately normalised background shapes from the sideband regions
in on-resonance data.
The dotted lines show the expected level of
\BB\ background, which is obtained from the sum of
Gaussian distributions from Monte Carlo estimated
cross-feed and $D^0 \pi$ events, each 
normalised to the number of events observed in on-resonance data 
that passed the selection criteria for $\modeV$.}
\label{B3piCut1K_dEandmESData}
\end{figure}

\begin{figure}[!htb]
\begin{center}
\hspace*{-1.0cm}
\mbox{\epsfig{file=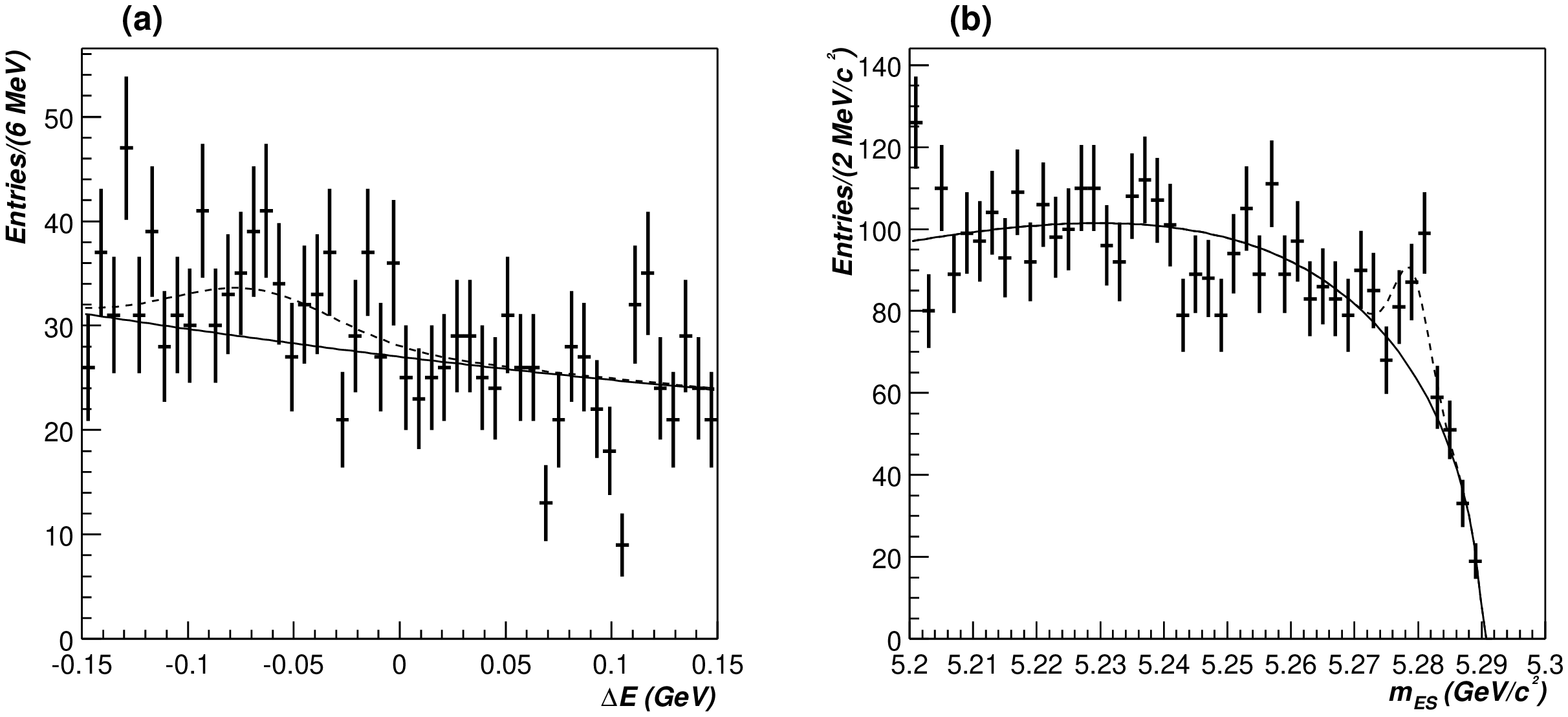,width=7.25cm,angle=-90}}
\end{center}
\caption[On-resonance signal region $\DeltaE$ and $\mes$ distributions for $\modeXI$]
{On-resonance signal region $\DeltaE$ (a) and $\mes$ (b) distributions
for $\modeXI$. The solid lines show the expected level of continuum background,
using appropriately normalised background shapes from the sideband regions
in on-resonance data.
The dotted lines show the expected level of
\BB\ background, which is obtained from the sum of
Gaussian distributions from Monte Carlo estimated cross-feed events, each 
normalised to the number of events observed in on-resonance data 
that passed the selection criteria for $\modeXI$.}
\label{B3piCut2K_dEandmESData}
\end{figure}

\begin{figure}[!htb]
\begin{center}
\hspace*{-1.0cm}
\mbox{\epsfig{file=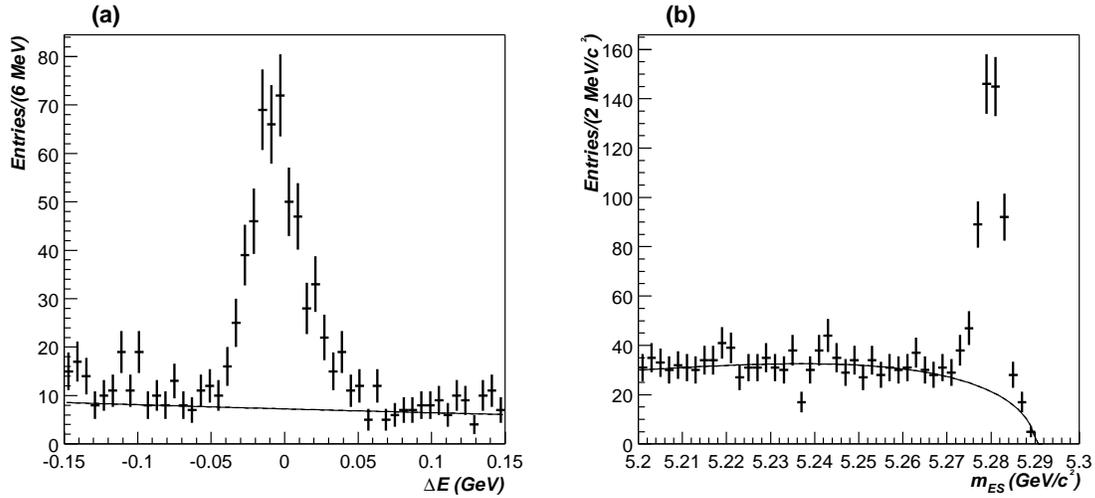,width=7.25cm,angle=-90}}
\end{center}
\caption[On-resonance signal region $\DeltaE$ and $\mes$ distributions for $\modeXII$]
{On-resonance signal region $\DeltaE$ (a) and $\mes$ (b) distributions
for $\modeXII$. The solid lines show the expected level of 
continuum background,
using appropriately normalised background shapes from the sideband regions
in on-resonance data.}
\label{B3piCut3K_dEandmESData}
\end{figure}

Figures~\ref{B3piCut0K_dEandmESData} to~\ref{B3piCut3K_dEandmESData} show the
$\DeltaE$ and $\mes$ distributions for the signal region for each of the modes.
Each plot shows the expected levels continuum and \BB\ background (solid and
dashed lines, respectively).

\begin{figure}[!htb]
\begin{center}
\hspace*{-1.0cm}
\mbox{\epsfig{file=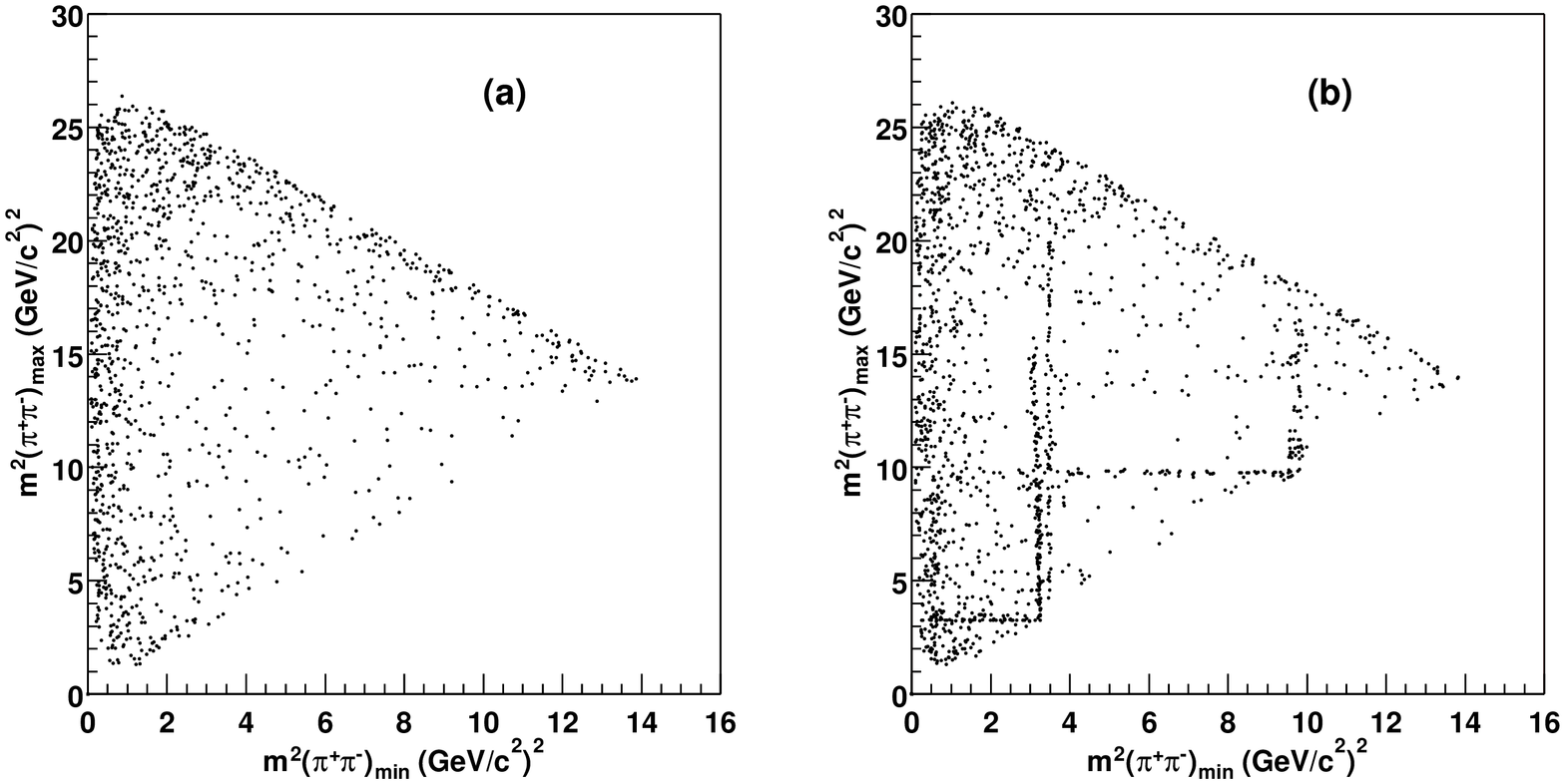,width=8.75cm,angle=-90}}
\end{center}
\caption[Dalitz plot for on-resonance data for $\modeIV$]
{Unbinned Dalitz plots for on-resonance data for $\modeIV$ for GSB region (a) 
and signal region (b). 
No efficiency corrections have been applied, and the open
charm contributions are included in the plots.}
\label{B3piDataDP}
\end{figure}

\begin{figure}[!htb]
\begin{center}
\hspace*{-1.0cm}
\mbox{\epsfig{file=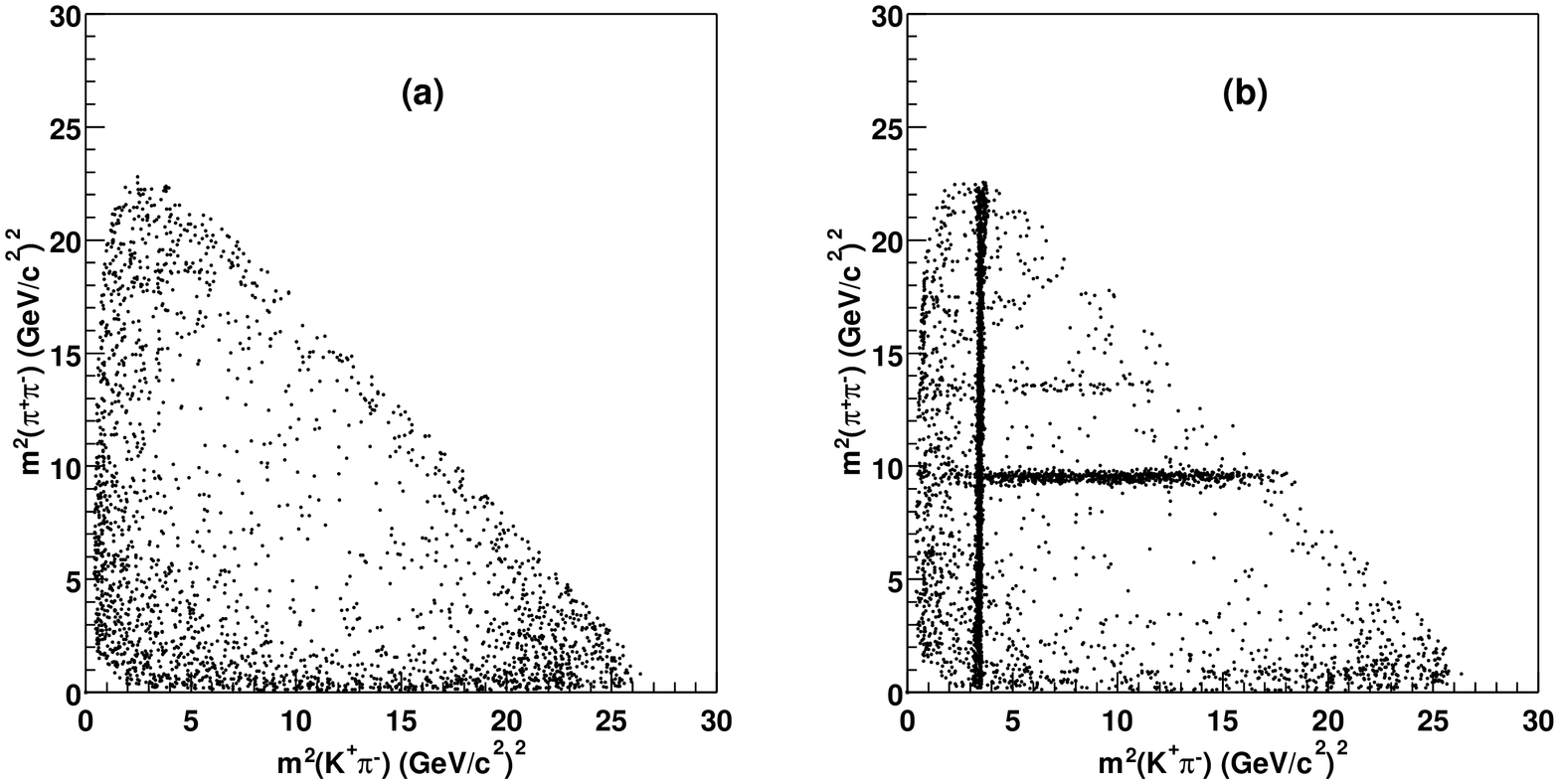,width=8.75cm,angle=-90}}
\end{center}
\caption[Dalitz plot for on-resonance data for $\modeV$]
{Unbinned Dalitz plots for on-resonance data for $\modeV$ for GSB region (a) 
and signal region (b). 
No efficiency corrections have been applied, and the open
charm contributions are included in the plots.}
\label{BKpipiDataDP}
\end{figure}

\begin{figure}[!htb]
\begin{center}
\hspace*{-1.0cm}
\mbox{\epsfig{file=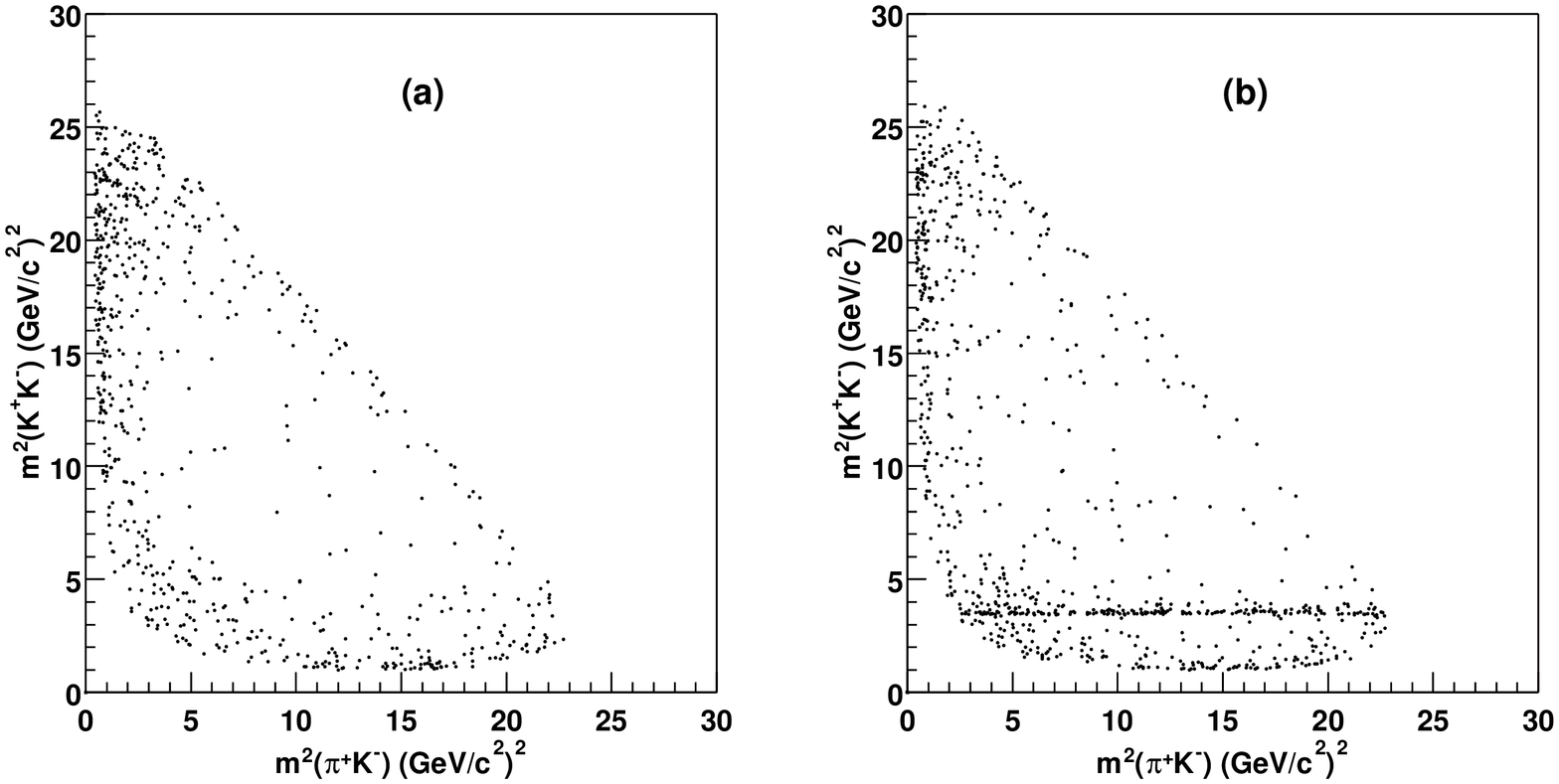,width=8.75cm,angle=-90}}
\end{center}
\caption[Dalitz plot for on-resonance data for $\modeXI$]
{Unbinned Dalitz plots for on-resonance data for $\modeXI$ for GSB region (a) 
and signal region (b). 
No efficiency corrections have been applied, and the open
charm contributions are included in the plots.}
\label{BKKpiDataDP}
\end{figure}

\begin{figure}[!htb]
\begin{center}
\hspace*{-1.0cm}
\mbox{\epsfig{file=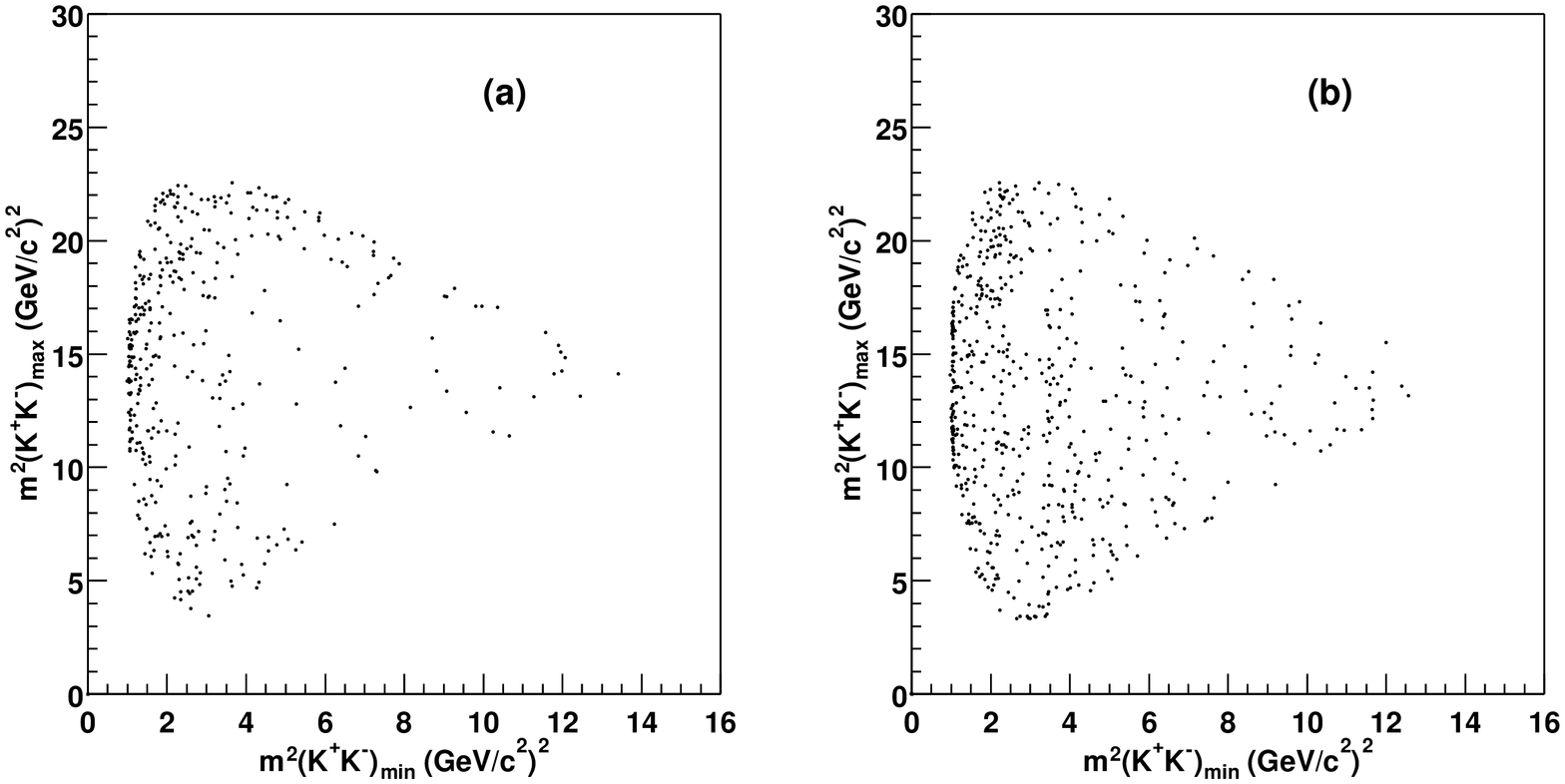,width=8.75cm,angle=-90}}
\end{center}
\caption[Dalitz plot for on-resonance data for $\modeXII$]
{Unbinned Dalitz plots for on-resonance data for $\modeXII$ for GSB region (a) 
and signal region (b). 
No efficiency corrections have been applied, and the open
charm contributions are included in the plots.}
\label{B3KDataDP}
\end{figure}

Figures~\ref{B3piDataDP} to~\ref{B3KDataDP} show the unbinned Dalitz plots for
the signal modes in the GSB and signal regions, where 
no efficiency corrections have been applied. 
Only the upper half of the symmetrical Dalitz plot
is shown for the $\modeIV$ and $\modeXII$ channels, where
the $x$ and $y$ axes show the
minimum and maximum values of the Dalitz plot variables, respectively.

There are clear signals for the modes $\modeV$ and $\modeXII$.
No signal is observed for $\modeXI$, and the result for $\modeIV$ is interpreted
as an upper limit on the branching fraction,
although there is a positive excess of signal events 
with $2.2~\sigma$ significance.
Since there are a large number of events in the selected samples, 
we can assume that 
the number of signal and background
events observed in the signal region are Gaussian distributed.
The 90\% C.L. upper limits are computed using the standard prescription
for a one-sided confidence interval from a Gaussian distributed
measurement, i.e.,
\begin{equation}
{\cal{B}}_{UL} = {\cal{B}} + 1.28 \Delta {\cal{B}},
\end{equation}
where ${\cal{B}}$ is the estimated branching ratio and
$\Delta{\cal{B}}$ is its standard deviation. Here, however, 
we take $\Delta{\cal{B}}$ to
be the total error (quadratic sum of statistical and systematic
uncertainties).

\begin{sidewaystable}[!htbp]
\caption{Branching fraction results for on-resonance data.
The uncertainties for each term are explained in the text.}
\label{tab:DataDPResults}
\begin{center}
\begin{tabular}{|l||c|c|c|c|}
\hline
& & & & \cr
Signal Mode & $\pi^{\pm} \pi^{\mp} \pi^{\pm}$ & $K^{\pm} \pi^{\mp} \pi^{\pm}$ 
& $K^{\pm} K^{\mp} \pi^{\pm}$ & $K^{\pm} K^{\mp} K^{\pm}$ \cr
& & & & \cr
\hline
\hline
& & & & \cr
No. of events in signal region, $\sum_i N_{1i}$ & 951 & 1269 & 573 & 603 \cr
\hline
& & & & \cr
No. of events in GSB, $\sum_i N_{2i}$ & 5470 & 4652 & 3239 & 1100 \cr
\hline
& & & & \cr
Average signal efficiency (\%) & $15.3 \pm 1.1$ & $15.4 \pm 0.9$ & 
$18.3 \pm 0.9$ & $22.5 \pm 1.0$\cr
\hline
& & & & \cr
Background factor, $R$ & $0.145 \pm 0.006$ & $0.153 \pm 0.006$ & $0.150 \pm 0.006$ & $0.159 \pm 0.010$ \cr
\hline
\hline
& & & & \cr
1) $\sum_i N_{1i}/\epsilon_i$ 
& $5839 \pm 212$ & $8055 \pm 255$
& $3414 \pm 156$ & $2734 \pm 111$\cr
\hline
& & & & \cr
2) $\sum_i RN_{2i}/\epsilon_i$ 
& $4812 \pm 73 \pm 193 $ & $4434 \pm 73 \pm 171$
& $2802 \pm 54 \pm 111$ & $780 \pm 23 \pm 47$\cr
\hline
& & & & \cr
3) $\sum_i N_x \epsilon''_i/\epsilon _i$ 
& $391 \pm 8 \pm 2$ & $14 \pm 1 \pm 1$ & $435 \pm 5 \pm 7$ & --- \cr
\hline
& & & & \cr
4) No. of $D^0$ events, $\sum_i n_{Di}/\epsilon_i$ & $157 \pm 27$ & $401 \pm 50$& --- & --- \cr
\hline
& & & & \cr
5) $2^{\mbox{nd}}$-order cross-feed, $n_x$ & --- & $-122 \pm 54$  & $57 \pm 11$ & --- \cr
\hline
& & & & \cr
6) {\LARGE $\sum_i \frac{(N_{1i} - R N_{2i} - N_x \epsilon''_i - n_{Di})} {\epsilon_i}$} $- n_x$
& $478 \pm 224 \pm 195$ & $3328 \pm 266 \pm 186$
& $120 \pm 166 \pm 112$ & $1954 \pm 114 \pm 47$ \cr
& $\pm 34 \pm 33$& $\pm 56 \pm 186$ & $\pm 22 \pm 6$ & $\pm 13 \pm 82$ \cr
& & & & \cr
\hline
\hline
& & & & \cr
Branching Fraction ($\times 10^{-6}$) 
& $8.5 \pm 4.0 \pm 3.6$ & $59.2 \pm 4.7 \pm 4.9$
& $2.1 \pm 2.9 \pm 2.0$ & $34.7 \pm 2.0 \pm 1.8$ \cr
Statistical Significance ($\sigma$)
& $2.2$ & $> 6$ & $0.9$ & $> 6$\cr
\hspace*{0.15cm} 90\% Upper Limit ($\times 10^{-6}$) 
& $< 15$ & & $< 7$ & \cr
& & & & \cr
\hline
\hline
\end{tabular}
\end{center}
\end{sidewaystable}


\clearpage


\section{Summary}
\label{sec:Summary}
We have obtained preliminary branching fractions for 
$\modeV$ and $\modeXII$ over the whole
Dalitz plot, and have determined conservative 90\% 
upper limits for $\modeIV$ and $\modeXI$. The results are summarised
in Table~\ref{tab:resCompare}, where the results from BELLE~\cite{ref:belle} 
are also included for comparison. 
\begin{table}[htb!]
\caption{Branching fraction results from \babar\ and BELLE.}
\begin{center}
{\small
\begin{tabular}{|l||c|c|}
\hline
Decay mode & \babar\ & BELLE \\
\hline
\hline
$\modepipipi$ & $< 15 \times 10^{-6}$ & --- \\
\hline
$\modeKpipi$ & $(59.2 \pm 4.7 \pm 4.9) \times 10^{-6}$ 
& $(58.5 \pm 7.1 \pm 8.8) \times 10^{-6}$\\
\hline
$\modeKKpi$ & $< 7 \times 10^{-6}$ & $< 21 \times 10^{-6}$\\
\hline
$\modeKKK$ & $(34.7 \pm 2.0 \pm 1.8) \times 10^{-6}$ 
& $(37.0 \pm 3.9 \pm 4.4) \times 10^{-6}$\\
\hline
\end{tabular}
}
\end{center}
\label{tab:resCompare}
\end{table}

\section{Acknowledgments}
\label{sec:Acknowledgments}

We are grateful for the 
extraordinary contributions of our \pep2\ colleagues in
achieving the excellent luminosity and machine conditions
that have made this work possible.
The success of this project also relies critically on the 
expertise and dedication of the computing organizations that 
support \babar.
The collaborating institutions wish to thank 
SLAC for its support and the kind hospitality extended to them. 
This work is supported by the
US Department of Energy
and National Science Foundation, the
Natural Sciences and Engineering Research Council (Canada),
Institute of High Energy Physics (China), the
Commissariat \`a l'Energie Atomique and
Institut National de Physique Nucl\'eaire et de Physique des Particules
(France), the
Bundesministerium f\"ur Bildung und Forschung
(Germany), the
Istituto Nazionale di Fisica Nucleare (Italy),
the Research Council of Norway, the
Ministry of Science and Technology of the Russian Federation, and the
Particle Physics and Astronomy Research Council (United Kingdom). 
Individuals have received support from 
the A. P. Sloan Foundation, 
the Research Corporation,
and the Alexander von Humboldt Foundation.

\end{document}